\documentclass[12pt]{iopart}

\usepackage{epsf,epsfig,psfig}

\def\spose#1{\hbox to 0pt{#1\hss}}
\def\lesssim{\mathrel{\spose{\lower 3pt\hbox{$\mathchar"218$}}
 \raise 2.0pt\hbox{$\mathchar"13C$}}}
\def\gtrsim{\mathrel{\spose{\lower 3pt\hbox{$\mathchar"218$}}
 \raise 2.0pt\hbox{$\mathchar"13E$}}}

\def\d{{\rm d}}
\def\tr{{\rm tr}\,}

\begin{document}

\title{Fixed point stability and decay of correlations}

\author{Ettore Vicari$\,^a$ and Jean Zinn-Justin$\,^{b,c}$}

\address{
$^a$ Dipartimento di Fisica dell'Universit\`a di Pisa and INFN, Pisa, Italy
}

\address{
$^b$ Dapnia, CEA/Saclay, F-91191 Gif-sur-Yvette cedex, France
}

\address{
$^c$ Institut de Math\'ematiques de Jussieu-Chevaleret, Universit\'e de Paris 7, France
}

\ead{Ettore.Vicari@df.unipi.it, zinn@dsm-mail.saclay.cea.fr}

\begin{abstract}
  In the framework of the renormalization-group theory of critical phenomena,
  a quantitative description of many continuous phase transitions can be
  obtained by considering an effective $\Phi^4$ theories, having an
  $N$-component fundamental field $\Phi_i$ and containing up to fourth-order
  powers of the field components.  Their renormalization-group flow is usually
  characterized by several fixed points. We give here strong arguments in
  favour of the following conjecture: the stable fixed point corresponds to
  the fastest decay of correlations, that is, is the one with the largest
  values of the critical exponent $\eta$ describing the power-law decay of the
  two-point function at criticality.  We prove this conjecture in the
  framework of the $\varepsilon$-expansion.  Then, we discuss its validity
  beyond the $\varepsilon$-expansion.  We present several lower-dimensional
  cases, mostly three-dimensional, which support the conjecture. We have been
  unable to find a counterexample.
\end{abstract}

\maketitle

\section{Introduction}
\label{intro}

In the framework of the renormalization-group (RG) approach to critical
phenomena, a quantitative description of many continuous phase transitions can
be obtained by considering effective Landau--Ginzburg--Wilson (LGW) $\Phi^4$
theories, having an $N$-component fundamental field $\Phi_i$ and containing up
to fourth-order powers of the field components.  The fourth-degree polynomial
form of the potential depends on the symmetry of the system.  LGW $\Phi^4$
theories generally present several fixed points (FP's) which are connected by
RG trajectories. See, for example, Fig.~\ref{omnflow}, which shows the RG flow
in the example of four fixed points. Among them, the infrared stable FP
determines the asymptotic critical behaviour of the corresponding statistical
systems. FPs are determined by the zeros of $\beta$ functions. The stability
of the FP is related to the eigenvalues of its stability matrix: if all
eigenvalues have a positive real part, then the FP is stable.  An interesting
question is whether a physical quantity exists such that the comparison of its
values at the FPs identifies the most stable FP.  In two dimensions, the
central charge is such a quantity: the stable FP in unitary theories is the
one with the least value of the central charge \cite{Zam-86}.  But, despite
several attempts and some progress, see, for example, Refs.~\cite{cc2}, no 
conclusive results on the extension of this
theorem to higher dimensions has been obtained yet.

In this paper we give strong arguments in favour of the following conjecture: 

{\em In general $\Phi^4$ theories with a single quadratic invariant, the
  infrared stable FP is the one that corresponds to the fastest decay of
  correlations.} 

Therefore, it is the FP with the largest value of the critical exponent $\eta$
which characterizes the power-law decay of the two-point correlation function
$W^{(2)}(x)$ at criticality,
\begin{equation}
W^{(2)}(x) \propto {1\over x^{d-2+\eta}}\,.
\label{etadef}
\end{equation}
The exponent $\eta$ is related to the RG dimension of the field, $d_\Phi=(d-2+\eta)/2$. 

The conjecture holds in the case of the $O(N)$-symmetric $\Phi^4$ theory.
Indeed, below four dimensions, the Gaussian FP, for which $\eta=0$, is
unstable against the non-trivial Wilson--Fisher FP for which $\eta\ge 0$.  We
recall that the positivity of $\eta$ in unitary theories follows rigorously
from the spectral representation of the two-point function \cite{ZJ-book}.

In the absence of a sufficiently large symmetry restricting the form of the
$\Phi^4$ potential, many quartic couplings must be introduced---see, for
example, Refs.~\cite{ZJ-book,BLZ-76,Aharony-76,PV-02,BLZ-74,variousphi4}.  The
Hamiltonian of a general $\Phi^4$ theory for a $N$-component field $\Phi_i$
can be written as
\begin{equation}
H = \int {\rm d}^d x \Bigl[ 
{1\over 2} \sum_i (\partial_\mu \Phi_{i})^2 + 
{1\over 2} \sum_i r_i \Phi_{i}^2  + 
{1\over 4!} \sum_{ijkl} u_{ijkl} \; \Phi_i\Phi_j\Phi_k\Phi_l \Bigr].
\label{genH}
\end{equation}
The number of independent parameters $r_i$ and $u_{ijkl}$ depends on
the symmetry group of the theory. An interesting class of models are
those in which $\sum_i \Phi^2_i$ is the unique quadratic polynomial
invariant under the symmetry group of the theory. In this
case, all $r_i$ are equal, $r_i = r$, and $u_{ijkl}$ must be such not
to generate other quadratic invariant terms under RG
transformations, for example, it must satisfy the trace
condition~\cite{BLZ-74} $\sum_i u_{iikl} \propto \delta_{kl}$.  In
these models, criticality is driven by tuning the single parameter
$r$, which physically may correspond to the temperature. All field
components become critical simultaneously and the two-point function
in the disordered phase is diagonal, that is,
\begin{equation}
W^{(2)}_{ij}(x-y) \equiv \langle 
\Phi_i(x) \Phi_j(y)\rangle = \delta_{ij} W^{(2)}(x-y).
\label{ediagonal}
\end{equation}
These $\Phi^4$ theories have generally several FPs.  Our conjecture applies to
such class of unitary models.  We do not consider nonunitary limits, such as
$N\rightarrow 0$, which are relevant to describe the critical properties of
spin systems in the presence of quenched disorder.

Actually, although within the $\varepsilon$-expansion around $d=4$ one can
prove that only one stable FP exists, in $d<4$ general $\Phi^4$ theories may
have more than one stable FP with separate attraction domains.  The $\eta$
conjecture should be then refined by comparing FP that are connected by RG
trajectories starting from the Gaussian FP: among them the stable FP is the
one with the largest value of $\eta$.

It was already observed in Ref.~\cite{BLZ-76} that, within the
$\varepsilon$-expansion, the stable FP of theories with four FPs, is the one
with the largest value of $\eta$.  Here, we extend this
$\varepsilon$-expansion result to an arbitrary $\Phi^4$ theory.  Then, we
discuss the validity of the $\eta$ conjecture at fixed dimension $d<4$, where
it remains a conjecture.  We present several lower-dimensional cases, mostly
three-dimensional, which support the conjecture. We have been unable to find
an analytical or numerical counterexample.

Finally, we extend the conjecture to multicritical points in models with
several independent correlation lengths (different $r_i$) that diverge
simultaneously. In this situation, the exponent $\eta$ is replaced by a matrix
and the conjecture applies to the trace of the matrix. However, the empirical
evidence, beyond the $\varepsilon$-expansion is more limited.

The paper is organized as follows. In Sec.~\ref{sec2-eps}, we prove
the $\eta$ conjecture within the $\varepsilon$-expansion for the most
general $\Phi^4$ with a single quadratic invariant.  In
Sec.~\ref{sec3-checks}, we discuss several lower dimensional examples,
showing that in all cases the stable FP is the one with the largest
value of $\eta$.  In Sec.~\ref{sec4}, we discuss the extension of the
$\eta$ conjecture to $\Phi^4$ theories describing multicritical
behaviours.  In Sec.~\ref{sec5} we discuss the Gross-Neveu-Yukawa
model, and show that also in this case the infrared stable FP is the
one with the fastest decay of the critical two-point function of the
boson field, within the $\varepsilon$-expansion and in the large-$N_f$
limit for any dimension.

\section{Proof within the $\varepsilon$-expansion}
\label{sec2-eps}

Expansions in powers of $\varepsilon=4-d$ can be most easily
obtained within the minimal-subtraction scheme \cite{tHV-72}, where the RG
functions are computed from the divergent part of correlation
functions \cite{ZJ-book},
\begin{eqnarray}
\beta_{ijkl}(g_{abcd})\equiv \mu {\partial g_{ijkl}\over \partial \mu},
\qquad 
\eta(g_{abcd})\equiv \mu {\partial {\rm ln} Z_\Phi \over \partial \mu},
\end{eqnarray}
where $g_{ijkl}$ are the renormalized couplings corresponding to the
quartic parameters $u_{ijkl}$.

We consider only Hamiltonians that have a symmetry such that the
quadratic invariant in the field is unique and the two-point function
in the disordered phase thus diagonal.  As a consequence of the
diagonal property of the two-point correlation function in the
disordered phase, the tensor $u_{ijkl}$ has special properties that
take the form of successive constraints in the perturbative expansion.  
At leading order one finds \cite{BLZ-74}
\begin{eqnarray}
\beta_{ijkl}(g_{abcd}) = 
-\varepsilon g_{ijkl}
+\frac{1}{16\pi^2} \sum_{m,n}\left(g_{ijmn}g_{mnkl}+g_{ikmn}g_{mnjl}+g_{ilmn}g_{mnkj}  \right) 
\end{eqnarray}
The RG function associated with the field dimension can be inferred from the function
\begin{equation}
\eta(g_{ijkl})   = 
{1\over 6N (4\pi)^4}\sum_{i,j,k,l} g_{ijkl} g_{ijkl}\,,
\label{etapert}
\end{equation}
this form resulting from the diagonality condition 
\begin{equation}
N \sum_{k,l,m}g_{iklm}g_{jklm}=
\delta_{ij}\sum_{ k,l,m,n}g_{klmn }g_{klmn}\,.
\end{equation}

One can easily verify that the general expression of the one-loop
$\beta$-function derives from a potential \cite{WZ-74}. Indeed,
\begin{eqnarray}
&&\beta_{ijkl}(g_{abcd})={\partial U(g_{abcd})\over\partial g_{ijkl}},
\label{potential} \\
&&
U(g_{abcd}) =-{\varepsilon\over2} \sum_{i,j,k,l}g_{ijkl}g_{ijkl}+
{1\over (4\pi)^2}\sum_{i,j,k,l,m,n}g_{ijkl}g_{klmn} g_{mnij} 
\nonumber 
\end{eqnarray}
Such a property, which has been verified to two-loop order, is shared, at
leading order, by other field theories and would deserve a more systematic
investigation.

A detailed discussion of the properties of the RG flow within the
$\varepsilon$-expansion can be found in Ref.~\cite{ZJ-book2}.  Here, we list a
number of consequences of Eq.~(\ref{potential}) within the
$\varepsilon$-expansion. Note that none of these properties depends on the
condition (\ref{ediagonal}).

(i) The potential decreases along a RG trajectory and thus FPs are extrema of
the potential. In particular, if two FPs are (asymptotically) connected by a
RG trajectory, the stablest FP corresponds to the lowest value of the
potential.

(ii) The eigenvalues of the matrix of first order partial derivatives of the
$\beta$ functions (stability matrix) at a FP are real.

(iii) Stable fixed points are local minima of the potential, that is, the
matrix of second derivatives of $U(g)$ is positive.

Moreover, two additional properties depend of the special cubic form of the
one-loop potential (\ref{potential}) (we give the proof in the appendix):

(iv) There exists at most one stable FP.

(v) The stable FP corresponds to the lowest value of the potential $U(g)$.

The latter properties are not necessarily valid beyond the
$\varepsilon$-expansion. For example, in the physical dimensions $d=3,2$ LGW
$\Phi^4$ theories may have more than one stable FPs with separate attraction
domains. This possibility does not occur within the $\varepsilon$-expansion,
that is, close to 4D, but it does not contradict general RG arguments and it
is found in some cases, in particular when different regions of the
Hamiltonian quartic parameters are related to different symmetry breaking
patterns.  Next section we shall mention a few examples where this occurs.

In the framework of the $\varepsilon$-expansion, we now show that the
stable FP (or at least the most stable one) corresponds to the largest
value of the exponent $\eta$, and thus to the case where correlation
function has the fastest decay at large distance.  For any
fixed point $g_{ijkl}^*$, the equations
\begin{equation}
\beta_{ijkl}(g_{abcd}^*) =  {\partial U(g_{abcd}^*)\over \partial g_{ijkl}}=0
\end{equation}
implies
\begin{equation}
\varepsilon \sum_{i,j,k,l} g^*_{ijkl}g^*_{ijkl}= 
{3\over(4\pi)^2}\sum_{i,j,k,l,m,n} g^*_{ijkl}g^*_{klmn}g^*_{mnij} 
\end{equation}
and, thus, at leading order,
\begin{equation}
U(g_{abcd}^*)=-\frac{1}{6}\varepsilon \sum_{i,j,k,l} g^*_{ijkl}g^*_{ijkl}
\end{equation}
which is negative, thus lower than the Gaussian FP value.
At leading order,
the exponent $\eta$ is then given by  
\begin{equation}
\eta={1\over6N}{1\over(4\pi)^4}\sum_{i,j,k,l}g^*_{ijkl}g^*_{ijkl}
=-{1\over N\varepsilon} {1\over(4\pi)^4}
U(g_{abcd}^*). 
\end{equation}
As we have shown, the stable fixed point corresponds to the lowest value of
$U$. It thus corresponds also to the largest value of the exponent $\eta$:
therefore, the correlation functions corresponding to the stable fixed point
have the fastest large distance decay.

The validity of this result beyond the $\varepsilon$-expansion remains a
conjecture. In next section, we discuss several checks in lower dimensions,
mostly in three dimensions.

\section{Several verifications of the $\eta$ conjecture}
\label{sec3-checks}

In this section, we discuss the RG flow, in $d<4$ dimensions, of several
$\Phi^4$ theories with a single quadratic invariant, but more than one quartic
terms.  As we shall see, in all examples considered below, the stable FP of
the RG flow, within regions connected by RG trajectories starting from the
Gaussian FP, is the one corresponding to the largest value of $\eta$.

\subsection{The $O(M)\otimes O(N)$ $\Phi^4$ model}
\label{omnsec}

We first consider the $O(M)\otimes O(N)$ $\Phi^4$ model corresponding to the
hamiltonian density
\begin{eqnarray}
{\cal H}&=&{1\over2}
      \sum_{a,i} [ (\partial_\mu \Phi_{ai})^2 + r \Phi_{ai}^2 ]
+ \frac{1}{4!} u_0 \left( \sum_{a,i} \Phi_{ai}^2 \right)^2 \label{omn}\\
&&+ \frac{1}{4!} v_0 \sum_{a,i,b,j} [\Phi_{ai}\Phi_{bi} \Phi_{aj}\Phi_{bj} 
- \Phi_{ai}^2 \Phi_{bj}^2]
\nonumber
\end{eqnarray}
where $\Phi_{ai}$ is a $M\times N$ real matrix
($a=1,...,M$ and $i=1,...,N$).
The symmetry of this model is $O(M)\otimes O(N)$.

\begin{figure}[tb]
\centerline{\psfig{width=6truecm,angle=0,file=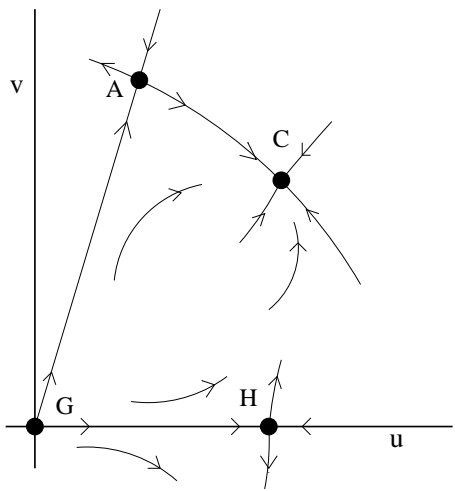}}
\vspace{2mm}
\caption{
RG flow of the $O(M)\otimes O(N)$ model in the large-$N$ limit.
}
\label{omnflow}
\end{figure}

\subsubsection{The large-$N$ limit}
\label{omnln}

The $O(M)\otimes O(N)$ $\Phi^4$ model can be solved in the large-$N$ limit for
any fixed $M$ \cite{Kawamura-91,PRV-01-ln}.  For any $M\ge 2$, one finds four
FP's: the Gaussian FP, the Heisenberg $O(M\times N)$ FP, and two new FP's
which we call chiral (C) and antichiral (A). Fig.~\ref{omnflow} shows a sketch
of the RG flow in the quartic-coupling space.  In the large-$N$ limit, for any
$M\ge 2$ and for any $2<d<4$, the stable FP is the chiral one, and all other
FP's are unstable.  The large-$N$ critical exponent $\eta$ has been calculated
for all FP's:
\begin{equation}
\eta = {\eta_1 e_d \over N} + O(1/N^2)
\end{equation}
where
\begin{equation}
e_d = - {4 \Gamma(d-2)\over \Gamma(2-d/2) \Gamma(d/2-1) 
\Gamma(d/2-2) \Gamma(d/2+1) }
\end{equation}
and
\begin{equation}
\eta_1= \, \cases{
    0 & \qquad\qquad {\rm Gaussian} \cr
    {1/M}   & \qquad\qquad  {\rm Heisenberg}\cr
    {(M+1)/2}  &  \qquad\qquad  {\rm chiral} \cr
    {(M-1)(M+2)/(2M)}  & \qquad\qquad  {\rm antichiral} 
    }
\label{eta1}
\end{equation}
$O(1/N^2)$ calculations can be found in Ref.~\cite{PRV-01-ln}.  One
easily verifies that for any $M$ and $2<d<4$ the value of $\eta$ at
the stable chiral FP is the largest.

\subsubsection{$d=3$ results from high-order FT perturbative analyses}
\label{omnm2}

Several results have also been obtained concerning the RG flow of the
three-dimensional $O(M)\otimes O(N)$ $\Phi^4$ models at finite values of
$M,N$.  Below we restrict ourselves to the case $M=2$, that is, to
$O(2)\otimes O(N)$ models.  The cases $N=2,3$ are physically interesting
because they could describe transitions in non-collinear frustrated magnets,
the superfluid transition in $^3$He, {\it etc}....  See, for example,
Refs.~\cite{Kawamura-98,PV-02} and references therein.  We have to distinguish
the cases $v_0>0$ and $v_0<0$, because they lead to different symmetry
breaking patterns:
\begin{eqnarray}
&&O(2) \otimes O(N) \rightarrow O(2) \otimes O(N-2) 
\quad {\rm for}\; v_0>0 \,,\label{v0p} \\
&&O(2)\otimes O(N) \rightarrow Z_2\otimes O(N-1)
\quad\quad {\rm for}\; v_0<0 \,.
\label{v0n}
\end{eqnarray}
The $u$-axis plays the role of a separatrix and thus the RG
flow corresponding to $v_0>0$ cannot cross the $u$-axis.  The relevant
FPs of models with the symmetry-breaking pattern (\ref{v0p}) lie in
the region $v>0$, where $v$ is the renormalized quartic coupling
associated with $v_0$, while the relevant FPs of models with the
symmetry-breaking pattern (\ref{v0n}) lie in the region $v\le 0$.

All $O(2)\otimes O(N)$ models contain the Gaussian FP and the
Heisenberg $O(2N)$ FP. They are both unstable.  The relevant
perturbation at the $O(2N)$ FP, which makes it unstable for any $N \ge
2$, is related to the $v$-term in the Hamiltonian,
which is a particular combination of quartic operators transforming as
the spin-0 and spin-4 representations of the O($2N$) group.  Any
spin-4 quartic perturbation is relevant at the O($K$) FP for $K\geq
3$, since its RG dimension $y_{4,4}$ is positive for $K\geq 3$
\cite{CPV-03}.  Therefore the $O(2N)$ FP is always unstable in the RG
flow of the $O(2)\otimes O(N)$ models (actually this result extends to
any $O(M)\otimes O(N)$ model with $M\ge 2$).  In particular,
$y_{4,4}\approx 0.11$ at the $O(4)$ FP and $y_{4,4}\approx 0.27$ at the
$O(6)$ FP.

The RG flow of three-dimensional $O(2)\otimes O(N)$ models has been
investigated by computing and analyzing high-order perturbative series within
the massive zero-momentum (MZM) and massless $\overline{\rm MS}$ schemes,
respectively to six and five loops \cite{PRV-01,CPS-02,DPV-04,CPPV-04}.  Some
results are reported below.

\begin{table}
\caption{\label{crexp}
Best theoretical estimates of the critical exponents for
three-dimensional $O(N)$ models. Concerning the methods, IHT indicates 
high-temperature expansion of improved lattice models with
suppressed leading scaling corrections; 
MC indicates Monte Carlo simulations;
FT indicates field-theoretical methods based on 
perturbative expansions;
MC+IHT exploits a synergy of IHT and MC.
Other results can be found in Refs.~\cite{ZJ-book,PV-02}
}
\begin{indented}
\item[]\begin{tabular}{@{}lllll}
\br
$N$ & $\nu$ & $\eta$ & Method& Ref. \\
\mr
1 & 0.63012(16) & 0.0364(2) & IHT & \cite{CPRV-02} \\
  & 0.63020(12) & 0.0368(2) & MC & \cite{DB-03} \\
  & 0.6304(13)  & 0.034(3)  & FT & \cite{GZ-98,LZ-77} \\
2 & 0.6717(1) & 0.0381(2)  & MC+IHT & \cite{CHPV-06} \\
  & 0.6703(15)  & 0.035(3)  & FT & \cite{GZ-98,LZ-77} \\
3 & 0.7112(5) & 0.0375(5)  & MC+IHT & \cite{CHPRV-02} \\
  & 0.7073(35)  & 0.0355(25)  & FT & \cite{GZ-98,LZ-77} \\
4  & 0.749(2) & 0.0365(10) & MC & \cite{Has-01} \\
   & 0.741(6) & 0.0350(45) & FT & \cite{GZ-98} \\
5  & 0.779(3) & 0.034(1)   & MC & \cite{HPV-05} \\
   & 0.764(4) & 0.031(3)   & FT & \cite{BP-05} \\
6  & 0.789(5) & 0.029(3)   & FT & \cite{BP-05} \\
8  & 0.830 & 0.027   & FT & \cite{AS-95} \\
large $N$ & $1-32/(3\pi^2 N)$   &  $8/(3 \pi^2 N)$ & $1/N$ exp & \cite{MZ-03} \\
\br
\end{tabular}
\end{indented}
\end{table}

{\em RG flow for $N=2$ and $v<0$.}
One finds a stable FP in the region $v<0$, which is in the $XY$
universality class \cite{Kawamura-98,PV-02}. We recall that the other
relevant FPs are the Gaussian FP and the $O(4)$ FP.  The
best available estimates of $\eta$ for $O(N)$ models are reported in
Table~\ref{crexp}. They support the $\eta$ conjecture, which would
require $\eta_{XY}>\eta_{O(4)}$.  Indeed the best available estimate
are $\eta_{XY}=0.0381(2)$ and $\eta_{O(4)} = 0.0365(10)$.

{\em RG flow for $N=2$ and $v>0$.}  The analysis of the high-order MZM
and $\overline{\rm MS}$ expansions provide a rather robust evidence of
the existence of another stable chiral FP for $v>0$
\cite{PRV-01,CPS-02,CPPV-04}. This has been confirmed by MC
simulations of a lattice $\Phi^4$ model
\cite{CPPV-04}.  This FP is not connected with the one found in 
the $v<0$ region, because the line $v=0$ is a separatrix.
The estimates of $\eta$ at this stable FP are:
$\eta_{ch}=0.09(1)$ from MZM and $\eta_{ch}=0.09(4)$ from
$\overline{\rm MS}$.   These results must be compared with the values of $\eta$ at 
the other FPs connected by RG trajectories in the region $v>0$,
which are the Gaussian and the $O(4)$ FPs.
Again the conjecture is verified because
$\eta_{ch}>\eta_{O(4)}>0$.

{\em RG flow for $N=3$ and $v<0$.}  There is a stable FP with
attraction domain in the region $v<0$ \cite{DPV-04}. The corresponding
estimates of $\eta$ are: $\eta=0.079(7)$ from MZM and $\eta=0.086(24)$
from $\overline{\rm MS}$.  The other FPs are the Gaussian FP and O(6)
FP, which have much smaller values of $\eta$, in particular
$\eta_{O(6)}=0.029(3)$ \cite{BP-05}.

{\em RG flow for $N=3$ and $v>0$.}  
There is a stable FP also in the region $v>0$ \cite{CPPV-04}.  The
corresponding estimates of $\eta$ are: $\eta=0.10(1)$ from MZM and
$\eta=0.08(3)$ from $\overline{\rm MS}$.  This values are again much
larger than the values of $\eta$ of the unstable Gaussian and $O(6)$
FPs.

Note that the stable three-dimensional FPs of $O(2)\otimes O(N)$ models with
$N=2,3$ do not exist close to four dimensions, apart from the one for $N=2$
and $v<0$, see, for example, Ref.~\cite{Kawamura-98}. Thus, they provide a
rather non-trivial check of the $\eta$ conjecture, because the
three-dimensional RG flow differs qualitatively from the RG flow close to
$d=4$, which is obtained from the $\varepsilon$-expansion.

\subsection{$\Phi^4$ theory with cubic anisotropy}
\label{seccubic}

\begin{figure}[tb]
\vspace{-2cm}
\centerline{\psfig{width=12truecm,angle=-90,file=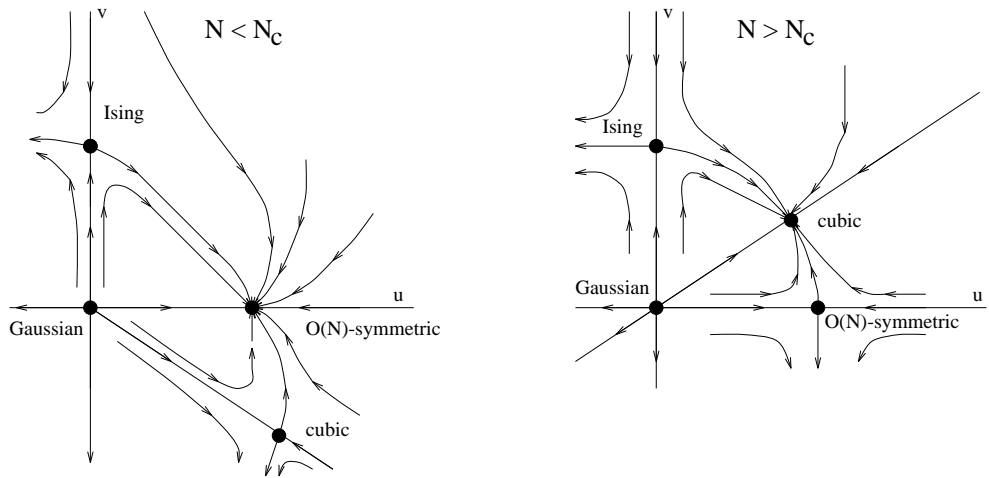}}
\vspace{-2cm}
\caption{
RG flow of the cubic model. $N_c\lesssim 3$ in $d=3$,
see, for example, Ref.~\cite{CPV-00}.
}
\label{cubicflow}
\end{figure}

This theory is relevant for magnets with cubic anisotropy.
Its Hamiltonian density is
\begin{eqnarray}
{\cal H} = \frac{1}{2} 
      \sum_i \left[ (\partial_\mu \Phi_i)^2 +  r \Phi_i^{2} \right]  
+ \frac{1}{4!} u_0 \left(\sum_i \Phi_i^2\right)^2  + \frac{1}{4!} v_0 \sum_i \Phi_i^4 ,
\nonumber
\end{eqnarray}
where $\Phi_i$ is a $N$-component field.  The RG flow of this
Hamiltonian has four FP's: the trivial Gaussian one, the Ising one in
which the $N$ components of the field decouple, the O($N$)-symmetric
and the cubic FP's.  The Gaussian FP is always unstable, and so is the
Ising FP for any number of components $N$.  Indeed, at the Ising FP
one may interpret the cubic Hamiltonian as the Hamiltonian of $N$
Ising systems coupled by the O($N$)-symmetric interaction.  The
coupling term $\int \d^d x \,\Phi_i^2\Phi_j^2$ with $i\neq j$ scales as
the integral of the product of two operators $\Phi_i^2$. Since the
$\Phi_i^2$ operator has RG dimension $1/\nu_I$---indeed, it is
associated with the temperature---the combined operator has RG
dimension $2/\nu_I - d = \alpha_I/\nu_I$, and therefore the associated
crossover exponent is given by $\phi=\alpha_I$, independently of $N$.
Since $\alpha_I>0$, the Ising FP is unstable independently of $N$.  On
the other hand, the stability properties of the O($N$)-symmetric and
of the cubic FP's depend on $N$.  For sufficiently small values of
$N$, $N<N_c$, the O($N$)-symmetric FP is stable and the cubic one is
unstable.  For $N>N_c$, the opposite is true: the RG flow is driven
toward the cubic FP, which now describes the generic critical
behaviour of the system.  Fig.~\ref{cubicflow} sketches the flow
diagram in the two cases $N<N_c$ and $N>N_c$.  In $d=3$ $N_c\approx
2.9$, see for example, \cite{CPV-00,PV-02} and references therein.  Therefore,
the $O(N)$ FP is stable only for $N=2$.
Let us now examine the various cases in more detail.

{\em The case $N=2$.}
We have four FPs: the Gaussian FP, the Ising FP along the $u=0$ axis, the
cubic FP for $v<0$ which turns out to be equivalent to an Ising FP, and the
$O(2)$ FP which is the stable one.  Again the $\eta$ conjecture does not fail:
it requires $\eta_{XY}>\eta_{\rm Ising}$ and this is verified by the best
three-dimensional estimates reported in Table~\ref{crexp}. i.e.
$\eta_{XY}=0.0381(2)$ and $\eta_{\rm Ising}=0.0364(2)$.

It is also worth mentioning the two-dimensional RG flow of this theory, where
there is a line of stable fixed points connecting the Ising and $XY$ FPs
\cite{JKKN-77,CC-02}, with central charge $c=1$ (in this case the central
charge associated with the Ising FP is $c=1$ because it represents two
decoupled Ising models).  Along this line the critical exponent $\eta$ does
not vary, $\eta=1/4$, but the correlation-length critical exponent goes from
$\nu=1$ to $\nu=\infty$.

{\em The case $N=3$.}  We have four FPs: the Gaussian FP, the Ising FP
along the $u=0$ axis, the $O(3)$ FP, and the cubic FP for $v>0$ which
should be the stable one.  The stable cubic FP turns out to be very
close to the $O(3)$ one, and the critical exponents are not
distinguishable with the best $O(3)$ estimates.  Indeed, FT estimates
of their differences give \cite{CPV-03-c}
$\nu_c-\nu_{O(3)}=-0.0003(3)$ and $\eta_c-\eta_{O(3)}=-0.0001(1)$.
The $\eta$ conjecture would require $\eta_c-\eta_{O(3)}>0$; the above
FT estimate of the difference $\eta_c-\eta_{O(3)}$, although favouring a negative sign, 
is not sufficiently precise to conclude that the $\eta$ conjecture fails.
In conclusion, also in this case the $\eta$
conjecture is substantially consistent with the available results: for $N=3$
$\eta_{\rm cubic}\approx \eta_{O(3)}>\eta_{\rm Ising}>0$.

{\em The case $N>3$.}
For $N>3$ the analysis of the six-loop series reported in
Ref.~\cite{CPV-00} is consistent with the conjecture. One finds
$\eta_{\rm cubic} \approx \eta_{\rm Ising}$, but the precision is not
sufficient to determine which one is larger.

{\em The case $N\rightarrow \infty$.}  
For $N\to \infty$, keeping $Nu$
and $v$ fixed, one can derive exact expressions for the exponents at
the cubic FP.  Indeed, for $N\to\infty$ the system can be
reinterpreted as a constrained Ising model, leading to a Fisher
renormalization \cite{Fisher-68}
of the Ising critical exponents.  One has
\begin{eqnarray}
\eta = \eta_I+O\left( {1/N}\right),\qquad
\nu = {\nu_I\over 1-\alpha_I}+O\left( {1/N}\right),
\label{largen}
\end{eqnarray}
where $\eta_I$, $\nu_I$, and $\alpha_I$ are the critical exponents of
the Ising model.  Again the $\eta$ conjecture does not fail.

\subsection{Spin-density-wave model}
\label{sdwsec}

The spin-density-wave model is a rather complicate $\Phi^4$ model with
five quartic parameters, which could describe the SDW-SC--to--SC
phase transition in high-$T_c$ superconductors (cuprates)
\cite{DPV-06}. Its Hamiltonian density is
\begin{eqnarray}
&&{\cal H}=\vert\partial_\mu \Phi_{1}\vert^2 +
                 \vert\partial_\mu \Phi_{2}\vert^2 
+r(\vert\Phi_{1}\vert^2+\vert\Phi_{2}\vert^2 )
+\frac{u_{1,0}}{2}(\vert\Phi_{1}\vert^4+\vert\Phi_{2}\vert^4)
\nonumber\\
&&+\frac{u_{2,0}}{2}(\vert\Phi_{1}^2\vert^2+\vert\Phi_{2}^2\vert^2)
   +w_{1,0}\vert\Phi_{1}\vert^2 \vert\Phi_{2}\vert^2
   +w_{2,0}\vert\Phi_{1}\cdot \Phi_{2}\vert^2 
   +w_{3,0}\vert\Phi_{1}^* \cdot \Phi_{2}\vert^2 
   \vphantom{\sum_\mu^d} 
\nonumber
\end{eqnarray}
where $\Phi_a$ are complex $N$-component vectors.  The RG flow of this
model has been investigated in Ref.~\cite{DPV-06}.  The physical
interesting cases are those for $N=2,3$. The analysis reported in
Ref.~\cite{DPV-06} suggests the existence of a stable FP in both
cases, with rather large values of $\eta$: $\eta=0.12(1)$ and
$\eta=0.18(3)$ respectively (from MZM). 
There are also several unstable FPs in the RG flow, but all of them 
have smaller values of $\eta$.  Therefore, the
available $d=3$ results of this spin-density-wave model support the
$\eta$ conjecture.

\subsection{$U(N)\otimes U(N)$-symmetric $\Phi^4$ models}
\label{u2ou2sec}

Let us now consider the Hamiltonian density
\begin{eqnarray}
{\cal H} = \tr (\partial_\mu \Phi^\dagger) (\partial_\mu \Phi)
+r \tr \Phi^\dagger \Phi 
+ {u_0\over 4} \left( \tr \Phi^\dagger \Phi \right)^2
+ {v_0\over 4} \tr \left( \Phi^\dagger \Phi \right)^2
\label{u2ou2}
\end{eqnarray}
where $\Phi_{ij}$ is a $N\times N$ complex matrix.  The symmetry is
$U(N)_L\otimes U(N)_R$.  In the case $\Phi_{ij}$ is also symmetric
the symmetry is $U(N)$. This model has been introduced and studied
\cite{PW-84,BPV-03-05} because it is relevant for the
finite-temperature transition in QCD, that is, the theory of the strong
interactions.  Indeed, for $v_0>0$ the ground state leads to the
symmetry-breaking pattern $U(N)_L\otimes U(N)_R\rightarrow U(N)_V$
(corresponding to QCD if the $U(1)_A$ anomaly is neglected).

For $N=1$, the model reduces to the $O(2)$-symmetric $\Phi^4$ theory.
For $N\ge 3$ no stable FP is found \cite{BPV-03-05}. 

We focus on the case $N=2$.  One can easily identify two FP's. One is the
Gaussian FP for $u=v=0$, which is always unstable.  Since for $v_0=0$ the
Hamiltonian becomes equivalent to the one of the $O(8)$-symmetric model, the
corresponding $O(8)$ FP must exist on the $v=0$ axis for $u>0$. The $O(8)$ FP
is also unstable because the $v$-term in the Hamiltonian represents a spin-4
perturbation with respect to the $O(8)$ FP, and such perturbations are
relevant for any $O(K)$ models with $K\ge 3$ \cite{CPV-03}. Within one-loop
$\varepsilon$-expansion calculations, no other FP is found \cite{PW-84}.  By
contrast, three-dimensional analysis of both MZM and $\overline{\rm MS}$
expansions show the presence of a stable FP in three dimensions
\cite{BPV-03-05}.  The corresponding value of $\eta$ is rather large:
$\eta\approx 0.1$, significantly larger than $\eta_{O(8)}\lesssim 0.03$.
Therefore, it supports the $\eta$ conjecture.

\subsection{$SU(4)$-symmetric $\Phi^4$ model}
\label{su4sec}

Let us now consider the Hamiltonian density
\begin{eqnarray}
{\cal H} &= \tr (\partial_\mu \Phi^\dagger) (\partial_\mu \Phi)
+r \tr \Phi^\dagger \Phi 
+ {u_0\over 4} \left( \tr \Phi^\dagger \Phi \right)^2
\nonumber \\
&+ {v_0\over 4} \tr \left( \Phi^\dagger \Phi \right)^2 
+ w_0 \left( \det  \Phi^\dagger +  \det  \Phi \right),
\label{su4}  
\end{eqnarray}
where $\Phi$ is a complex and symmetric $4\times 4$ matrix field.
The symmetry of this model is $SU(4)$.
For $v_0 > - {3\over2} |w_0|$, the theory describes
the symmetry-breaking pattern $SU(4)\rightarrow SO(4)$ 
\cite{BPV-03-05}, which is the appropriate symmetry-breaking pattern 
to describe transitions in a QCD-like theory  
with quarks in the adjoint representation.

Close to four dimensions there are only two FPs, the Gaussian and the
$O(20)$ FPs, which are both unstable.  They remain unstable even at
lower dimensions and thus are of no relevance for the critical
behaviour. The three-dimensional RG flow has been investigated by
field-theoretical methods based on perturbative approaches, within the
MZM and $\overline{\rm MS}$ schemes \cite{BPV-03-05}.  They show the
presence of a stable three-dimensional fixed point characterized by
the symmetry-breaking pattern ${\rm SU}(4)\rightarrow {\rm SO}(4)$.
The corresponding value of $\eta$ is rather larger: $\eta\approx 0.2$,
much larger than the values of $\eta$ of the Gaussian and $O(20)$ FPs
which is $\eta_{O(20)}\approx 0.013$.

\section{$\Phi^4$ theories of multicritical behaviours}
\label{sec4}

In this section, we discuss the extension of the $\eta$ conjecture to $\Phi^4$
theories describing multicritical behaviours, characterized by more than one
independent correlation lengths. In this situation, the $\varepsilon$-expansion indicates
that the stable FP should be the one with the largest value of the trace of
the $\eta$ matrix.

We discuss this point within the $\Phi^4$ theory 
\begin{eqnarray}
{\cal H}=(\partial_\mu \vec\phi_1)^2 
+(\partial_\mu \vec\phi_2)^2+ r_1 \vec\phi_1^{\,2}+ r_2 \vec\phi_2^{\,2}
+ u_1 (\vec\phi_1^{\,2})^2 + u_2 (\vec\phi_2^{\,2})^2 + 
w \vec\phi_1^{\,2} \vec\phi_2^{\,2} 
\end{eqnarray}
where $\phi_{1,2}$ are two O($n_1$) and O($n_2$) order parameters, with $n_1$
and $N_2$ real components respectively.  The symmetry is $O(n_1)\oplus
O(n_2)$. This $\Phi^4$ theory describes the multicritical behaviour arising
from the competition of orderings with symmetries O($n_1$) and O$(n_2)$, at
the point where the corresponding transition lines meet in the phase diagram
\cite{FN-74}. Such multicritical points arise in several physical contexts,
for instance in anisotropic antiferromagnets, in high-$T_c$ superconductors, 
etc.... See, for example, Ref.~\cite{CPV-03} and references therein.

The multicritical behaviour is determined by the RG flow in the
quartic-coupling space when $r_{1,2}$ are tuned to their critical values.
Four FPs are found: the Gaussian FP, the isotropic
$O(n_1+n_2)$ FP (describing an effective enlargement of the symmetry), a
decoupled $O(n_1)$-$O(n_2)$ FP (which describes effectively decoupled order
parameters), and a biconal FP. The main properties of the
three-dimensional RG flow are the following.

(i) For $n_1+n_2\ge 4$ the decoupled FP is stable.  
This can be inferred from non-perturbative arguments \cite{Aharony-02} 
that show that the RG dimension $y_w$
of the perturbation $P_w=\phi_1^2 \phi_2^2$ that couples the two order
parameters is 
\begin{eqnarray}
y_w =  {1\over \nu_{1}} + {1 \over \nu_{2}} - 3,
\end{eqnarray} 
where $\nu_1,\nu_2$ are the correlation-length exponents of the
O($n_1$) and O($n_2$) models.  Inserting the numbers reported in Table~\ref{crexp}
one finds $y_w<0$
for $n_1+n_2\ge 4$ and any $n_1,n_2\ge 1$. 

(ii) Field-theoretical methods based on perturbative expansions (six-loop in
the MZM and $O(\varepsilon^5)$ in the $\varepsilon$-expansion) show that the
isotropic $O(n_1+n_2)$ FP is unstable for $n_1+n_2\ge 3$ \cite{CPV-03}.
Therefore, only in the case of two Ising order parameters can the symmetry 
be effectively enlarged from ${\rm Z}_2\oplus {\rm
  Z}_2$ to $O(2)$, at the multicritical point where the Ising lines meet.

(iii) For $n_1=1$ (Ising), $n_2=2$ ($XY$), $O(\varepsilon^5)$ calculations show
that the stable FP is the biconal FP \cite{CPV-03}.  Its critical exponents
turn out to be very close to the $O(3)$ ones, in fact they are not
distinguishable within the errors of the best estimates of the
$O(3)$ critical exponents, see Table \ref{crexp}.

If we compare the isotropic and the decoupled FPs, the conjecture on the
trace of the $\eta$ matrix should give
\begin{equation}
 n_1 \eta_{O(n_1)} + n_2 \eta_{O(n_2)} > (n_1+n_2) \eta_{O(n_1+n_2)} 
\end{equation}
for $n_1+n_2\ge 4$, and
\begin{equation}
\eta_{XY} > \eta_{\rm Ising}
\end{equation}
Moreover, from the point (iii) above, we should have
\begin{equation}
\tr \eta_{\rm biconal} > \eta_{\rm Ising} + 2 \eta_{XY}
\end{equation}
All these relations are verified by, or when the precision is not sufficient
are consistent with, the best estimates of the exponent $\eta$ for the $O(N)$ models,
see Table~\ref{crexp}.

\section{The Gross--Neveu--Yukawa model}
\label{sec5}

In this final section we discuss the Gross--Neveu--Yukawa (GNY) model \cite{ZJfer},
 and show that the infrared stable
FP of its RG flow is the one characterized by the fastest decay of the critical
two-point function of the boson field.
 
The Lagrangian of the GNY model is 
\begin{equation}
{\cal L} = -\sum_i \bar{\psi}_i( \gamma_\mu \partial_\mu + g_0 \sigma)\psi_i + 
\frac{1}{2}  [ (\partial_\mu \sigma)^2 + m^2 \sigma^2]
+ \frac{1}{4!} u_0 \sigma^4, 
\label{gnymodel}
\end{equation}
where $\psi_i$ are $N_f$ fermionic fields, 
and $\sigma$ is a real scalar field.  The relation
between the GNY and the standard Gross--Neveu model is discussed in
Ref.~\cite{ZJfer}.

Its RG flow can be investigated within the $\varepsilon$-expansion. 
The RG functions at one-loop order are \cite{ZJfer}
\begin{eqnarray}
&&\beta_u = - \varepsilon u + \frac{1}{8\pi^2} \left( \frac{3}{2} u^2 + N u g^2 - 6
  N g^4\right), \label{betasGNY}\\
&&\beta_{g^2} = - \varepsilon g^2 + \frac{N+6}{16\pi^2} g^4,
\nonumber
\end{eqnarray}
where $N=N_f \tr I$ (the trace is in the $\gamma$-matrix space) is the
total number of fermion components.  In four dimensions $\tr I=4$, and
thus $N=4 N_f$ in Eq.~(\ref{betasGNY}).  $u$ and $g$ are the ${\rm MS}$
renormalized couplings associated with $u_0$ and $g_0$, respectively. Note that at one-loop the RG $\beta$ function for a completely general Yukawa coupling \cite{ColGro} derives also for a potential (see also \cite{ZJ-book}).\par
The $\beta$ functions of the GNY model have three FPs.  Beside the unstable
Gaussian FP, there is an Ising FP at
\begin{equation}
u_* = {16\pi^2 \over 3} \varepsilon,\qquad g^2_*=0,
\label{isinggny}
\end{equation}
which is also unstable. The infrared stable FP of the theory is the
Gross--Neveu FP at
\begin{equation}
u_* = {384 N\pi^2 \over (N+6) [(N-6) + (N^2+132N+36)^{1/2}]}
\varepsilon,\qquad g^2_*= {16\pi^2\over N+6}\varepsilon\,.
\label{gngny}
\end{equation}
We can now compare the corresponding RG dimensions of the scalar
field, and check if the infrared stable FP is the one with the largest
value of $\eta_\sigma$.  Calculations of the scalar field RG functions
show that the critical exponent $\eta_\sigma$ is maximum at the GN FP:
\begin{equation}
\eta_\sigma = {N\over N+6}\varepsilon\,,
\end{equation}
while $\eta_\sigma=O(\varepsilon^2)$ at the Ising FP.

The GNY model is soluble in the large-$N_f$ limit for any $d$ \cite{ZJfer}, for a review see Ref.~\cite{MZ-03}.  In
this limit one finds
\begin{equation}
\eta_\sigma = 4-d + O(1/N_f),
\label{largeNlimit}
\end{equation}
and therefore $\eta_\sigma=1+O(1/N_f)$ in three dimensions.  The values of 
$\eta_\sigma$ at the Ising FP are definitely smaller, for example
$\eta_\sigma\approx 0.036$ in three dimensions, see Table~\ref{crexp}, and
$\eta_\sigma=1/4$ in two dimensions.

In conclusions, the above analytical results suggest that in the GNY
model the infrared stable FP is the one that corresponds to the
fastest decay of correlations of the scalar field, like in the
$\Phi^4$ theories.

\appendix

\section{Some proofs within the $\varepsilon$-expansion}

In the framework of the $\varepsilon$-expansion, we prove two consequences of
the property of gradient flow discussed in Sec.~\ref{sec2-eps}: (i) there
exists at most one stable fixed point; (ii) the stable fixed point corresponds
to the lowest value of the potential.  Indeed, let us assume the existence of
two fixed points corresponding to the parameters $g ^*$ and $g'{}^*$.  We then
consider the parameters $g$ of the form
\begin{equation}
g(s)=s g ^*+(1-s)g'{}^*\,, \quad 0\le s\le 1,
\end{equation}
and the corresponding potential $u(s)=U\bigl(g(s)\bigr)$.
As the explicit form (\ref{potential}) shows, at leading order  $u(s)$  
is a third degree polynomial in $s$.
The derivative
\begin{eqnarray}
u'(s)=\sum_a g'_a(s){\partial U \over \partial g_a}
=\sum_a (g^*_a-g_a'{}^*){\partial U \over \partial g_a} =\sum_{a}
(g^*_a-g_a'{}^*) \beta _a\big(g(s)\bigr) 
\end{eqnarray}
vanishes due to the fixed point conditions at $s=0$ and $s=1$:
$u'(0)=u'(1)=0$.
Since $u'(s)$ is a second degree polynomial, it then has necessarily the form
\begin{equation}
u'(s)=  As(1-s).
\end{equation}
The second derivative  $u''(s)$ is given in terms of the matrix of second partial
derivatives of $U$ and, thus, the partial derivatives of the $\beta $-functions, by
\begin{equation}
u''(s)=\sum_{a,b}(g_a^*-g_a'{}^*) 
{\partial^2 U \bigl(g(s)\bigr)\over\partial g_a\partial g_b}(g_b^*-g_b'{}^*)
=A(1-2s)
\end{equation}
In particular, for $s=0$  and $s=1$
\begin{eqnarray}
A &=&\sum_{a,b}(g_a ^*-g_a'{}^*)  {\partial^2 U(g'{} ^*  ) \over\partial
  g_a\partial g_b}    (g _b^*-g_b'{}^*), \\
-A &=&\sum_{a,b}(g_a ^*-g_a'{}^*)   {\partial^2 U(g  ^*  ) \over\partial
  g_a\partial g_b}   (g _b^*-g_b'{}^*). 
\end{eqnarray}
At a stable fixed point, the matrix ${\bf U} '' $ of partial second
derivatives  of $U$ is positive. Thus, if $g^*$  and $g'{}^*$ 
are stable fixed points, $A$  and $-A$ are both given by the expectation
 value of a positive matrix and thus are both positive, 
which is contradictory: the two fixed points cannot both be stable.

More generally, the sign of $A$ characterizes, in some sense, the
relative stability of these two fixed points. Let us assume, for example,
$A <0$ which is consistent with the assumption that $g^*$ is
stable. Then $u'(s)<0$ in $[0,1]$ and $U (g(s) )$ is a
decreasing function. Thus,
\begin{equation}
U(g^*)<U(g'{}^*).
\end{equation}
In particular, if $g ^*$ is a stable fixed point, it corresponds,
among all fixed points, to the lowest value of the potential.

\end{document}